\documentclass[showpacs,aps]{revtex4}
\usepackage{graphicx}

\begin{document}

\title{A conjecture on the infrared structure of the vacuum Schr\"odinger
wave functional of QCD}
\author{John M. Cornwall\footnote{Email:  Cornwall@physics.ucla.edu}}
\affiliation{Department of Physics and Astronomy, University of California,
Los
Angeles CA 90095
\begin{center}
{\rm (Received February xx, 2007)}
\end{center}}

\begin{abstract}
\pacs{11.15.-q, 12.38.-t, 11.15.Tk   \hfill UCLA/06/TEP/29}

The Schr\"odinger wave functional $\psi=\exp-S\{\mathcal{A}^a_i(\vec{x})\}$
for the $d=3+1$ QCD vacuum is a partition function constructed in $d=4$;  
the exponent $2S$ [in $|\psi|^2=\exp (-2S)$] plays the role of a $d=3$
Euclidean action.  We start from a simple  conjecture for $S$ based on
dynamical generation of a gluon 
mass $M$ in $d=4$, then use earlier techniques of the author to extend (in
principle) the
conjectured form to full non-Abelian gauge invariance.   We argue that the
exact leading term, of
$\mathcal{O}(M)$, in an expansion  of $S$ in inverse powers of $M$ is a
$d=3$ gauge-invariant mass term (gauged non-linear sigma model); the next
leading term, of $\mathcal{O}(1/M)$, is a conventional Yang-Mills action.
The $d=3$
action that is (twice) the sum of these two terms has center vortices as
classical
solutions.  The $d=3$  gluon mass $m_3$, which we constrain to be the same
as $M$, and $d=3$ coupling $g_3^2$ are related through the conjecture to
the $d=4$ coupling strength, but at the same time the dimensionless ratio
$m_3/g_3^2$ can be  estimated from $d=3$ dynamics.  This allows us to
estimate the $d=4$ coupling $\alpha_s(M^2)$ in terms of the strictly
$d=3$ ratio $m_3/g_3^2$; we find a value of about 0.4, in good agreement
with an earlier theoretical value but somewhat low compared to the QCD
phenomenological value of $0.7\pm 0.3$. The wave functional for $d=2+1$ QCD
has an exponent
that is a $d=2$ infrared-effective action having both the
gauge-invariant mass term and the field strength squared term, and  so
differs from the conventional QCD action in two dimensions, which has no
mass term.  This conventional $d=2$ QCD would lead in $d=3$  to confinement
of all color-group representations.  But with the mass term (again leading
to center vortices), only $N$-ality $\not\equiv 0$ mod $N$
representations can be confined (for gauge group $SU(N)$), as expected. 
\end{abstract}

\maketitle

\section{\label{intro} Introduction}

The functional Schr\"odinger equation (FSE) for gauge theories, while no
simpler to solve (and perhaps harder, in some ways) than any other
non-perturbative formulation of QCD, has often been used over the years to
gain insight into various aspects of QCD or, more generally, $SU(N)$ gauge
theory
\cite{loos,green79,halpern,green80,jackiw80,feynman,ja84,corn87,corn92,
mansfield94,kk,rein97,zar,pachos,mansfield99,feurein,greenole}.  However,
few of these works address the important question of how confinement is
expressed in the FSE.  

In any approach to the FSE for QCD that purports to reveal confinement,
there are
two important prerequisites:  The first is gauge invariance, and it has
been addressed many ways.  The second is the need to insure that there are
only short-range field-strength correlations; otherwise (see, {\em e.~g.},
the qualitative and in some ways incomplete discussion of Feynman
\cite{feynman}) there cannot be confinement.  Given these, confinement
further requires long-range   pure-gauge contributions to the potential. 
These long-range
pure-gauge parts appear in the FSE as massless longitudinally-coupled
scalars that mimic Goldstone fields, although of course there is no
symmetry breaking in QCD.  Just as with conventional Goldstone fields, these
massless poles do not appear in the QCD S-matrix; this would be so even if
QCD were not a confining theory.  As is well-known, center vortices,
solitons of an infrared-effective action for QCD that
encapsulates dynamical and gauge-invariant generation \cite{corn82,papaag}
of a gluon mass $M$, show just these properties and so provide a
confinement mechanism.  This mass has been estimated theoretically
\cite{corn82}, from phenomenology \cite{field,natale}, and on the lattice
\cite{deforc}, all yielding values of 600$\pm$200 MeV.  The center vortices
in $d=3$ are
characterized by closed strings that (generically) constitute the
constant-time cross-sections of $d=4$ center vortices; a confining
condensate of center vortices in $d=4$ is therefore mirrored by a similar
condensate in $d=3$.
(Of course, the classical local minimum describing a
single or a few center vortices is not relevant in isolation; it is
necessary that there be a entropy-driven condensate of vortices.  We do not
discuss that issue here.)

In $d=3+1$ the FSE describes four-dimensional dynamics in $d=3$ terms,
because the exponent $S$ in the vacuum wave functional $\psi = \exp (-S)$
is (half of) a $d=3$ action,  which we label $I_{d=3}$, depending on the
spatial gauge potentials at
zero time.  Many authors have discussed center vortices for QCD in strictly
$d=4$ terms.     Our question is, how are
such solitons---and hence confinement---described in the FSE action
$2S=I_{d=3}$?

Our answer proceeds in four steps.  The FSE exponent $S$ is an infinite
series of $n$-point functions integrated over the spatial components of $n$
gauge potentials (see   the Appendix, which
reviews earlier work \cite{corn87} on the FSE, as well as
Sec.~\ref{nonabelian}).  The first step, described in Sec.~\ref{abelian},
considers the lowest-order term $S_2$ of this expansion, which is quadratic
and shows only Abelian $U(1)^{N^2-1}$ gauge invariance.   Our conjectured
form
of $S_2$ exactly satisfies the FSE with an Abelian gauge Hamiltonian that
phenomenologically describes a gauge-invariant gluon mass $M$; it
is essentially $N^2-1$ copies of the Abelian Higgs model with infinite Higgs
mass.    

Since our focus is on confinement, an infrared phenomenon, we will use
techniques and approximations that are useful in the infrared regime, even
though they may misstate ultraviolet-dominated phenomena.  In particular,
although we treat the gluon mass $M$ as a constant, it is actually a
running mass $M(k^2)$ evaluated on-shell.  In order that there be dynamical
mass generation in QCD, the running mass must vanish for large momentum
$k^2$ \cite{corn82}.  This vanishing cures certain short-distance
singularities of the center-vortex solitons coming from an
infrared-effective action.  We will ignore this complication throughout
this paper.

The Abelian case is
not entirely trivial, since the action $S_2$ contains the square root of an
operator---the hallmark of the FSE. (Throughout this paper, we take this
operator, called $\Omega$, in the simple form $\Omega
=\sqrt{M^2-\nabla^2}$.)  Nonetheless, $S_2$ has center-vortex solutions. 
Although these do not completely coincide with conventional $d=3$ center
vortices, they show the necessary features:  Long-range pure-gauge parts
that confine, and field strengths that vanish at large distances as $\exp
(-M\rho )$ where $\rho$ is the distance from the  closed string on which
the vortex lives.  

In anticipation of what we must do in the non-Abelian case, we study briefly
the infrared expansion of $S_2$ in powers of $k^2/M^2$, and show that the
first two terms yield a familiar action.  The leading term is a
gauge-invariant mass term; the next-leading term is the usual Abelian gauge
action.  However, if the expansion is truncated after two terms, the gauge
mass described by them is erroneous.  The reason is elementary:  The
infrared expansion, at least in the Abelian case, is nothing but the first
two terms of the expansion
\begin{equation}
\label{2termexp}
\sqrt{M^2-\nabla^2}\rightarrow \frac{1}{M}(M^2-\frac{1}{2}\nabla^2 +\dots );
\end{equation}
the two terms saved correspond to a mass $\sqrt{2}M$ instead of $M$.
We propose that it may be phenomenologically useful, although not highly
accurate, to make the replacement
\begin{equation}
\label{meansq}
\sqrt{M^2-\nabla^2}\rightarrow \frac{Z}{M}(M^2-\nabla^2)
\end{equation}
where the renormalization constant $Z\simeq 1$ can be estimated in various
ways.  This heuristic replacement has the correct gluon mass.  We discuss
the motivation for this renormalization, coming from omitted terms in the
infrared expansion.

The second step, the
subject of Sec.~\ref{nonabelian}, begins with the problems of  enforcing
non-Abelian gauge
invariance.  Using earlier work \cite{corn87}, we show that $S_2$ of step
one can be gauge-completed to exact non-Abelian gauge invariance with an
infinite series of $n$-point functions and powers of gauge potentials, in
such a way that all $n$-point functions depend only on the operator
$\Omega$, no matter what the specific form of $\Omega$ is.   Gauge
completion uses the pinch technique \cite{corn82,binpap} and the gauge
technique \cite{cornhou}, as reviewed in the Appendix.  The gauge technique
is an
approximation that becomes exact only at zero momentum, but is useful
generally for momenta not large compared to $M$.  In this gauge-completed
$S$ we continue to use the simple form in $S_2$ of the two-point function
introduced in the Abelian case.  Just as for the ordinary Schr\"odinger
equation, either direct substitution in the FSE or a dressed-loop expansion
based on \cite{cjt} ultimately yields a non-linear Schwinger-Dyson equation
for $\Omega$ (see the Appendix).  We do not attempt to carry out this
difficult program to find $\Omega$, but simply use the form already
introduced in the Abelian case, showing mass generation.

Even the approximate (although showing full non-Abelian gauge invariance)
form of $S$ coming from application of the gauge/pinch technique is
extremely complex, involving not only square roots of operators but an
infinity of terms.  This $S$ looks nothing like actions that we are used to
dealing with.  Ultimately, whatever form $S$ takes must be dealt with on
its own terms.  However, just as in the Abelian case it can be helpful to
look for an approximate but familiar form.  We make the same sort of mass
expansion, saving only the first two terms, and argue that for QCD the
leading term, of
$\mathcal{O}(M)$, is equivalent to a gauged non-linear sigma (GNLS) model,
which
is commonly used as a description of gauge-invariant dynamical mass
generation in Yang-Mills theory (see, for example, \cite{corn74,corn82}). 
This sigma model contains the massless scalar poles, actually pure-gauge
parts of center vortices, that are responsible for confinement.  The second
term, of $\mathcal{O}(1/M)$, is (after gauge completion) the conventional
Yang-Mills action.  But as in the Abelian case, the mass is wrong by a
factor $\sqrt{2}$, so we suggest using the replacement of
Eq.~(\ref{meansq}).

In Sec.~\ref{massexpand} we give the final conjecture for the non-Abelian
exponent $S$ and $d=3$ action $I_{d=3}\equiv 2S$, and the main consequences
following from it. The conjectured action is the sum of a GNLS model and a
conventional
Yang-Mills action, with the correct free-field mass and a poorly-known
renormalization constant $Z$.  We suggest a method or two for estimating
$Z$, probably with no more than 25\% accuracy.

The fourth step is to examine the consequences of this final two-term
action.  We have already noted that this action has center vortices as
classical local minima (classical maxima of the FSE wave functional), and
thus could
provide a description of confinement in the FSE, which was one of our
principal goals.   Moreover, by appealing to known $d=3$ gauge
dynamics, we
can estimate the $d=4$ coupling strength in terms of the renormalization
constant $Z$.  In $d=3$ the coupling $g_3^2$
has dimensions of mass, and there is a unique [for given $SU(N)$]
dynamically-determined ratio $M/g_3^2$, which has been estimated by a
number of authors
\cite{chk,cornyan,alexn,buchphil,corn1998,eber,karsch,hkr,ckp,naka,nair}. 
Knowing only this ratio we can estimate the $d=4$ QCD
coupling $\alpha_s(M^2)$, getting a value around 0.4$Z$.  For
$Z\simeq 1$ this is
reasonably close both to an early $d=4$ estimate \cite{corn82} using the
gauge technique and pinch
technique but somewhat low compared to phenomenological estimates
\cite{natale} of
0.7$\pm$0.3.  Another application is to the $d=2+1$ FSE, studied in, among
other works, \cite{green79,greenole}.  Our present techniques suggest that
the corresponding $d=2$ FSE exponent $S$ is again a sum of a
gauge-invariant mass term and the usual Yang-Mills action.  Greensite
\cite{green79} speculated that this $S$ just had the conventional
Yang-Mills term.  However, as noted there and in \cite{greenole}, this
would lead to the wrong conclusion that in $d=2+1$ all  representations of
$SU(N)$ were confined, when in fact the adjoint and other representations
with $N$-ality $\equiv 0$ mod $N$ are screened, not confined.  But with the
addition of the mass term, confinement can  come about through center
vortices, and this form of confinement correctly predicts screening for
these representations.   

The paper ends with Sec.~\ref{conclusions}, giving conclusions.  An Appendix
reviews some background material on the FSE, including applications of the
pinch/gauge technique to the gauge FSE.

\section{\label{abelian} Describing mass generation in the FSE:  The Abelian
case}

Notation:  Throughout this paper we will always use the
canonical gauge potential $\mathcal{A}_i^a(\vec{x})$ potential
multiplied by the coupling $g$, with the notation:
\begin{equation}
\label{pot2}
A_i^a(\vec{x})=g\mathcal{A}_i^a(\vec{x}).
\end{equation}
Here $a$ is a group index for gauge
group $SU(N)$, and $i=1,2,3$ index the spatial components. All vectors are
three-dimensional, so we will now drop the vector notation and just use,
{\em e.g.,} $k$ for a three-momentum.
We also use the antihermitean matrix form
\begin{equation}
\label{pot}
A_i(x)=(\frac{g}{2i})\lambda_a\mathcal{A}^a_i(x) 
\end{equation}
where the $\lambda_a$ are
the Gell-Mann matrices for $SU(N)$, obeying
\begin{equation}
\label{trace}
Tr\frac{1}{2}\lambda_a\frac{1}{2}\lambda_b=\frac{1}{2}\delta_{ab}.
\end{equation}
The $A_i^a$ have engineering mass dimension 1 in any dimension.   The time
component $A_i^0$ is missing from the FSE. In this paper we will not need
to indicate gauge-fixing and ghost terms necessary to define the $d=3$
functional integrals that yield physical expectation values.

In the first step we begin with a simple quadratic (in the gauge potentials)
form for $S$ that is
consistent with gluon mass generation.  This quadratic form $S_2$ is
Abelian,
showing $U(1)^{N^2-1}$ local gauge invariance:
\begin{equation}
\label{prelim}
S_2=\frac{1}{2g^2}\int A_i^a\Omega_{ij}A_j^a(x)
\end{equation}
where the integral is over three-space, and $\Omega_{ij}$ is a product of
two factors:
\begin{equation}
\label{omegadef}
\Omega_{ij}=P_{ij}\Omega.
\end{equation} 
The factor $P_{ij}$ is a transverse projector: 
\begin{equation}
\label{pijdef}
P_{ij}=\delta_{ij}-\frac{\partial_i\partial_j}{\nabla^2}
\end{equation}
that is required for Abelian gauge invariance.
The free-field value of $\Omega$, called $\Omega_0$, describes free massless
particles:
\begin{equation}
\label{freeomega}
\Omega_0=\sqrt{-\nabla^2}=\sqrt{k^2}
\end{equation}
where $k$ is a three-momentum.  To describe dynamical mass generation we
will use, in this paper, the simple form
 \begin{equation}
\label{massomega}
\Omega = \sqrt{-\nabla^2+M^2}
\end{equation}
in which the gluon mass $M$ is the on-shell value of a running mass. 
Putting
these equations together we have:
\begin{equation}
\label{s2form}
S_2=\frac{1}{2g^2}\int A_i^a\sqrt{M^2-\nabla^2}P_{ij}A_j^a.
\end{equation}

One can easily check that   $S_2$  
is an exact solution to the FSE for an Abelian Hamiltonian with a
gauge-invariant mass term:
\begin{equation}
\label{abelham}
H=\int\{-\frac{1}{2}g^2(\frac{\delta}{\delta A_i^a})^2+
\frac{1}{2g^2}[\frac{1}{2}(F_{ij}^a)^2+M^2A_i^aP_{ij}A_j^a]\}\equiv 
\int [\frac{1}{2}(\Pi_i^a)^2]+V.
\end{equation}
where $F_{ij}^a=\partial_iA_j^a-\partial_jA_i^a$ are the Abelian field
strengths.
Here the mass term is put in by hand; in the non-Abelian version, we imagine
that this mass term summarizes the effects of non-Abelian condensates.

\subsection{Equations of motion and solitons for $S_2$}

One goal in this Abelian example is to find center vortex-like solitons as
extrema of $S_2$.    It may not be entirely obvious how to proceed, because
this action has the square root of an operator, leads to subtleties
concerning positivity, locality, and
self-adjointness.  For example, we will see that the operator
$\sqrt{M^2-\nabla^2}$ effectively vanishes on center vortex solitons,
although $-\nabla^2$ is  formally positive; this would falsely suggest that
the action of such a soliton is zero.    Consider the following alternative
description of $S_2$, found by expanding the square root in powers of
$-\nabla^2/M^2$ and assuming that integration by parts with no boundary
terms is allowed at all orders:
\begin{equation}
\label{adjointsqrt}
S_2=\frac{M}{2g^2}\int A_i^aP_{ij}A_j^a+\frac{1}{4g^2}\int
\sum_{N=0}C_{N+1}M^{-1-2N}
  [\partial_1\dots \partial_N F_{ij}^a(x)]^2
\end{equation} 
where $\partial_k\equiv \partial /\partial x_k$ and the   $C_N$ are the
 coefficients of $x^N$ in the power-series expansion of $\sqrt{1+x}$.  This
re-definition of the square root gives  the same generalized Euler-Lagrange
equations as the naive equations following from the original form of
Eqs.~(\ref{prelim},\ref{omegadef},\ref{massomega}), because these equations
assume that
integrating by parts gives no contributions (as would be appropriate for
functions  that fall off at least exponentially).

In order to study these generalized Euler-Lagrange equations, it is very
helpful to have $S_2$ in a formally local form.   We note that, term by
term, all but the first term of this alternative form of $S_2$ are both
local and manifestly gauge-invariant, and need no change.  As for the first
term, we replace (in a familiar way) the non-local part by scalar
fields:
\begin{equation}
\label{localize}
S_2  =  \frac{M}{2g^2}\int [A_i^a-\partial_i\phi^a]^2
 +\frac{1}{8Mg^2}\int [F_{ij}^a]^2
+\dots 
\end{equation}
Now keeping only a finite number of terms in the mass expansion of $S_2$
yields a local action, although of course the infinite sum may introduce
non-localities.

Saving only the first two terms in the mass expansion of $S_2$ based on
Eq.~(\ref{adjointsqrt})  should fail to satisfy the Abelian FSE based on
the Hamiltonian of Eq.~(\ref{abelham}).  It is instructive to work out this
failure and its consequences.   
The FSE reads:
\begin{equation}
\label{fse}
\frac{-g^2}{2}\int (\frac{\delta S_2}{\delta A_i^a})^2
+\frac{g^2}{2}\int \frac{\delta^2S_2}{\delta A_i^a \delta A_i^a}
+H=E
\end{equation}
where $E$ is the vacuum energy.  Since the second-derivative term on the
left-hand side of this equation only contributes to $E$, we drop it and
renormalize $E$ to zero.  
The mass expansion of $S_2$ suggests that the remaining quadratic term in
the FSE is in error at $\mathcal{O}(1/M^2)$.  A simple calculation confirms
this; Eq.~(\ref{fse}) becomes:
\begin{equation}
\label{fseerror}
\frac{-g^2}{2}\int (\frac{\delta S_2}{\delta A_i^a})^2+H +
\frac{1}{4g^2}\int \frac{1}{2M^2}(\partial_jF_{ij}^a)^2=0.
\end{equation}
 
At least qualitatively this error term in the FSE [last term on the
left-hand side, of $\mathcal{O}(1/M^2)$, the same relative order as the
$N=1$ term in Eq.~(\ref{adjointsqrt})] can
be thought of as increasing the kinetic  field-strength term $(F_{ij}^a)^2$
by a factor involving a mean-square momentum of the type $\langle
k^2\rangle/M^2$; such an increase helps restore the balance between kinetic
and mass terms in the expanded Hamiltonian which was disrupted by the usual
infrared expansion of Eq.~(\ref{localize}).  Such a renormalization is not
quantitatively trivial, since momenta relevant for solitons such as center
vortices are of $\mathcal{O}(M)$.

It is useful to restate the local form of $S_2$ in a compact way, by undoing
the power-series expansion and integration by parts:
\begin{equation}
\label{resum}
S_2 = \frac{M}{2g^2}\int [A_i^a-\partial_i\phi^a]^2 +\frac{1}{2g^2}
\int A_i^aP_{ij}[\sqrt{M^2-\nabla^2}-M]A_j^a.
\end{equation}
The scalar fields $\phi^a$ are to be integrated over, which may be thought
of as projection of a simple mass term $(A_i^a)^2$ onto its gauge-invariant
part by integrating over all gauge transformations.  Because the $\phi^a$
appear quadratically, such an integration is the same as solving the
classical field equations.  The field equations for the $\phi^a$ are
identical with a constraint following from the field equations for the
$A_i^a$.

Varying $S_2$, one finds the gauge potential equations of motion:
\begin{equation}
\label{abelfeqn}
M(A_i^a-\partial_i\phi^a)+[\sqrt{M^2-\nabla^2}-M]P_{ij}A_j^a=0.
\end{equation}
The divergence yields the $\phi^a$ equations:
\begin{equation}
\label{phieqn}
\nabla^2\phi^a=\partial_iA_i^a\rightarrow \phi^a=\frac{1}{\nabla^2}
\partial_iA_i^a+\varphi^a\;\;{\rm with}\;\;\nabla^2\varphi^a =0.
\end{equation}
Re-write  Equation (\ref{phieqn}) as:
\begin{equation}
\label{rewrite}
\sqrt{M^2-\nabla^2}P_{ij}A_i^a=M\partial_i(\phi^a-\frac{1}{\nabla^2}
\partial_jA_j^a)=M\partial_i\varphi^a.
\end{equation}
Multiplication by $\sqrt{M^2-\nabla^2}$ leads to:
\begin{eqnarray}
\label{sqrtmult}
(M^2-\nabla^2)P_{ij}A_j^a & = &
M\sqrt{M^2-\nabla^2}\partial_i\varphi^a\rightarrow
\\ \nonumber
\nabla^2A_i^a-\partial_i\partial_jA_j^a & = & M^2(A_i^a-\partial_i\frac{1}
{\nabla^2}\partial_jA_j^a)-M\sqrt{M^2-\nabla^2}\partial_i\varphi^a 
\rightarrow \\ \nonumber
\nabla^2A_i^a-\partial_i\partial_jA_j^a -
M^2(A_i^a-\partial_i\phi^a) & = &
M[M-\sqrt{M^2-\nabla^2}]\partial_i\varphi^a. \\ \nonumber
\end{eqnarray}
 
Term by term, every term on the right-hand side of the third equation in
Eq.~(\ref{sqrtmult}) vanishes, if we use $\nabla^2\varphi^a=0$. Since in
$\mathcal{R}^3$ there are no fields $\varphi^a$ solving
$\nabla^2\varphi^a=0$ that are regular everywhere and vanish at infinity, 
one may be tempted to make the stronger statement that $\varphi^a$ must
vanish. But the description of center vortices requires a non-zero
$\varphi^a$, singular on a closed Dirac hypersurface of
co-dimension 2 (a closed string in $d=3$), so it is more accurate to
say that term by term the expansion of the right-hand side of the third
equation in Eq.~(\ref{sqrtmult})  vanishes almost everywhere.  However, we
will soon see that this is not true for the unexpanded form.
If we nonetheless drop this term with the square-root operator,  the final
equations of motion are the usual equations \cite{corn79} for center
vortices, the same
as would be gotten from the $d=3$ Euclidean action
\begin{equation}
\label{d3action}
\frac{1}{2g^2}\int
\{M^2(A_i^a-\partial_i\phi^a)^2+\frac{1}{2}(F_{ij}^a)^2\}.
\end{equation}
  This action is just the
potential $V$ occurring in the Abelian Hamiltonian of Eq.~(\ref{abelham}),
written in local form; it is the Abelian form of the $d=3$
infrared-effective action used \cite{corn79,corn82} to describe mass
generation, and it has center vortices as classical solitons.  

If the term  $M[M-\sqrt{M^2-\nabla^2}]\partial_i\varphi^a$ is left
unexpanded, things are slightly different, although there are still center
vortices characterized by long-range pure-gauge parts and field strengths
vanishing exponentially as $\exp (-M\rho)$, where $\rho$ is the distance
from the Dirac string.  A center vortex is always fully determined by
$\varphi^a$.  We present our results in the gauge $\partial_iA_i^a=0$, in
which case $\phi^a =\varphi^a$.  The well-known expression \cite{corn79} for
the
center-vortex $\partial_i\phi^a$ is:
\begin{equation}
\label{phiform}
\partial_i\phi^a(x)=2\pi Q^a \epsilon_{ijk}\partial_j\oint_{\Gamma} dz_k\int
\frac{d^3k}
{(2\pi )^3}\frac{1}{k^2}e^{ik\cdot (x-z)}
\end{equation}
where the closed contour $\Gamma$ is the Dirac string, and $Q^a$ is one of
the $N-1$ generators of the Cartan subalgebra, normalized so that
 $\exp (2\pi i Q)$ is in the center of $SU(N)$.  Now the third equation in
Eq.~(\ref{sqrtmult}) easily gives:
\begin{equation}
\label{solve3d}
A_i^a(x)=2\pi Q^a\epsilon_{ijk}\partial_j\oint_{\Gamma}dx_k
\int \frac{d^3k}
{(2\pi )^3}\frac{M}{k^2\sqrt{k^2+M^2}}e^{ik\cdot (x-z)}.
\end{equation}
In the usual $d=3$ vortex, an extremum of the action in
Eq.~(\ref{d3action}), the factor $M(k^2+M^2)^{-1/2}$ would be replaced by 
$M^2(k^2+M^2)^{-1}$.

This unusual square root does not change the fact that the field strengths
show exponential decrease; in fact:
\begin{equation}
\label{bfield}
B_i^a=\frac{1}{2}\epsilon_{ijk}F_{jk}^a=2\pi Q^a\oint_{\Gamma}dz_i
\frac{M^2}{2\pi^2|x-z|}K_1(M|x-z|).
\end{equation}
There is, of course, still the long-range pure-gauge part associated with
$\phi^a$, which we can isolate by the decomposition:
\begin{equation}
\label{decompose}
\frac{M}{k^2\sqrt{k^2+M^2}}=\frac{1}{k^2}+
\frac{1}{k^2}\{\frac{M}{\sqrt{k^2+M^2}}-1\}.
\end{equation}
The second term on the right-hand side is short-ranged.
The short distance behavior is more singular than that of the conventional
vortex, but leads only to a logarithmic singularity in the value of $S_2$,
the same as for the conventional vortex.  In both cases the singularity is
multiplied by a power of $M$, which removes the singularity because the
running mass vanishes at short distances.
So the vortex extrema of $S_2$ differ in detail from the usual center
vortex, but have the hallmark features of a long-range pure-gauge part and
field strengths vanishing like $\exp (-M\rho )$.

\subsection{Mass expansion of $S_2$}

Another goal of this section is to replace $S_2$, which is either  given in
Eq.~(\ref{adjointsqrt}) as an infinite sum involving derivatives of
arbitrarily high order or in Eq.~(\ref{resum}) in terms of square roots of
operators,  by a tractable and recognizable action.   The first two terms
of the  expansion, written explicitly in Eq.~(\ref{localize}, fit these
criteria, but suffer from a serious defect.  The coefficient of the second
term, the usual gauge action, is wrong by a factor of 2; as written, it
describes gauge bosons of mass $\sqrt{2}M$.  This wrong coefficient arises
from the expansion $\sqrt{1+x}=1+(x/2)+\dots $.  We can see the same thing
happening with a mass expansion of the Fourier kernel of
Eq.~(\ref{decompose}).  Expand the square root in the  curly brackets of
this equation in powers of $k^2/M^2$ to get:
\begin{equation}
\label{fourexp}
\frac{M}{k^2\sqrt{k^2+M^2}}=\frac{1}{k^2}-\frac{1}{k^2+2M^2}+\dots
\end{equation}
This is exactly the kernel of the usual $d=3$ vortex, but with the wrong
mass $\sqrt{2}M$.  This is not the only way of expanding; for example,
re-writing the Fourier kernel in a different form and expanding the square
root occurring in it gives:
\begin{equation}
\label{reexpand}
\frac{M}{k^2\sqrt{k^2+M^2}}=\frac{M\sqrt{k^2+M^2}}{k^2(k^2+M^2)}=
\frac{M^2}{k^2(k^2+M^2)}-\frac{1}{2(k^2+M^2)}+\dots
\end{equation}
The first term on the right-hand side is the standard center vortex with the
correct mass $M$, and all other terms have this mass as well.  However,
these other terms give the wrong coefficient for the exponential falloff of
the field strengths at large distance.  

There are no such results for square-root operators in the non-Abelian case,
which is as expected much more complicated.  So we will, in the spirit of
the Abelian expansion given in Eq.~(\ref{localize}), look for a way to
approximate the complicated non-Abelian result by a two-term form, the
first of which is a (gauge-invariant) mass term and the second is the usual
Yang-Mills action.  In the Abelian case, such a two-term action as an
approximation to the infinite sum of Eq.~(\ref{adjointsqrt}) suggests that
the derivatives in this sum, beyond those in $F_{ij}^a$, be approximated by
averages so that this equation is effectively
\begin{equation}
\label{s2eff}
S_2=\frac{M}{2g^2}\int A_i^aP_{ij}A_j^a+\frac{1}{4Mg^2}\int
\sum_{N=0}C_{N+1}\langle \frac{k^{2N}}{M^{2N}}\rangle
  [F_{ij}^a(x)]^2\equiv \frac{M}{2g^2}\int
A_i^aP_{ij}A_j^a+\frac{Z}{4Mg^2}\int [F_{ij}^a(x)]^2
\end{equation}
where $k^{2N}$ stands for the multiple derivatives.  If this is justified,
the infinitely many terms of Eq.~(\ref{adjointsqrt}) are indeed replaceable
by a mass term plus a renormalized conventional gauge action.    But because
the gluonic mass described by this $S_2$ must be $M$, the same as in the
original $S_2$, there will have to be an equal renormalization of the mass
term in Eq.~(\ref{s2eff}) above. 
In later sections we will explore an approximation to the square root that
is motivated by these remarks, involving the replacement
\begin{equation}
\label{replace}
\sqrt{M^2-\nabla^2}\rightarrow \frac{Z}{M}(M^2-\nabla^2)
\end{equation}
for some renormalization constant $Z$, supposed to be near unity.  We have
no reliable techniques for calculating $Z$, so we will resort to a
simplistic approach of making a least-squares fit of the operator
$\sqrt{M^2-\nabla^2}$ by the operator $(Z/M)(M^2-\nabla^2)$, which leads to
$Z\simeq 1.1-1.2$.

Before engaging in
this mass expansion we must understand the gauge structure of the
non-Abelian exponent $S$.

\section{\label{nonabelian} The non-Abelian case:  Gauge completion and mass
expansion}   

 We can be nowhere near as complete in the non-Abelian case as we were
above, and ultimately will be forced to resort to a large-mass expansion.   

In the non-Abelian case, the
quadratic term $S_2$ with which we began is supplemented with an infinity of
terms, involving
spatial integrals over $n\geq 3$ spatial gauge potentials multiplied by an
$n$-point function $\Omega_n$ depending on the spatial and discrete
coordinates of the gauge potentials (see the Appendix):   
\begin{equation}
\label{nonabel1}
g^2S=\frac{1}{2!}\int \int A_i^a\Omega_{ij}A_j^a+\frac{1}{3!}\int \int \int
A_i^aA_j^bA_k^c
\Omega_{ijk}^{abc}+\dots
\end{equation}
The $n$-point function of this
expansion is related to the $n+1$-point function through ghost-free Ward
identities, as arise in the pinch technique \cite{corn82,binpap}.  These
Ward identities can be ``solved" using the gauge technique, a well-known
technique whose main points of interest we describe in the Appendix, and
the result is that it is possible in principle to find an approximate but
exactly gauge-invariant expression for the entire series of $n$-point
functions describing the wave functional exponent $S$.  Each $n$-point
function depends only on the two-point function, but in a complicated way
that is not understood.  Ultimately the two-point function is determined by
a non-linear Schwinger-Dyson equation that can (again in principle) be
derived either by direct substitution in the FSE or by a dressed-loop
expansion \cite{cjt,corn98,ccds}.  Using a
dressed-loop expansion  for $S$ is equivalent
to direct solution of the FSE (of course, either the dressed-loop expansion
or the FSE must be truncated at at certain number of loops, but this
truncation has nothing to do with a truncation in the coupling $g^2$;
all-order non-perturbative effects arise even at one-dressed-loop order in
QCD).  

A systematic study of the FSE would go on to determine the mass $M$ from the
infinity of equations for the $n$-point functions in $S$ of
Eq.~(\ref{nonabel1}), but that is not our purpose here.   Instead, we show
how to construct what we will call a {\em gauge completion} of the 2-point
action for an arbitrary $\Omega$, using earlier work \cite{corn87}, to add
higher-point functions, consistent with solving the FSE, that depend on
$\Omega$ in specific ways that insure full
non-Abelian gauge invariance.  Ultimately, the FSE becomes a non-linear
equation for $\Omega$, just as for the ordinary Schr\"odinger equation.

Gauge invariance
requires that the lowest-order (quadratic) term has the Abelian form already
given in Eq.~(\ref{s2form}).  The Ward identities for the three-point
function and their solution are detailed in the Appendix.  Both the Ward
identities and the FSE for the determination of this three-point function
involve only the two-point function $\Omega_{ij}$, and it is plausible that
there exists a three-point function satisfying these equations that is a
functional solely of the two-point function $\Omega_{ij}$.  The gauge
technique provides such a three-point function, as given in
Eqs.~(\ref{revfse},\ref{cornhoueq}).  The gauge technique by itself does
not furnish a unique solution, which must be found by recourse either to
the FSE itself or to the dressed-loop expansion.  However, in the infrared
limit of momenta small compared to the mass $M$ the solution is unique.

\subsection{Mass expansion:  The leading term}

In general, the gauge/pinch technique leads to quite complicated
expressions, and  we will explore only a simplified version of it.  The
main simplification is to look at the leading terms in an expansion in
inverse powers of $M$.  In the leading term, of $\mathcal{O}(M)$, all
two-point functions $\Omega$ are replaced just by $M$ itself, which gets
rid of many momentum-dependent terms.  In this way the leading term of the
three-point function is:
\begin{equation}
\label{3ptfunct}
\Omega_{ijk}^{abc}(k_1,k_2,k_3)=f^{abc}\frac{M}{6}\{\frac{k_{1i}k_{2j}(k_1-k_2)_k}
{k_1^2k_2^2}+c.p.\}+\mathcal{O}(1/M).
\end{equation}

One can proceed in principle this way, by looking at the pinch/gauge
technique solution for the four-point function (see \cite{papa}) and taking
the large-mass limit, then the five-point function, etc.  We will not
detail such an investigation here, but will point out some features that
strongly suggest the all-order solution.  The structure of the Ward
identities shows that the leading term of every $n$-point function is
$\mathcal{O}(M)$, with all other dimensions taken up by momenta, and that
the gauge-technique solution involves  longitudinally-coupled massless
poles whose number grows with $n$.  Observe further that the GNLS term of
$S$ is the exact solution of an FSE Hamiltonian consisting of just  this
term itself, as given in Eq.~(\ref{gnlsm}) below, multiplied by $M$.  Of
course, there is no such term in the underlying QCD Hamiltonian, but there
would be one in the infrared-effective Hamiltonian of QCD, derived by $d=4$
techniques \cite{corn74,corn82}.  

We suggest that the action of the gauged non-linear GNLS model,
expressed in non-local form [as in the originally-stated form of $S_2$, in
Eq.~(\ref{s2form})]   is the all-order perturbative solution to the leading
mass terms
of the gauge/pinch technique approach.   To find this non-local form we
investigate the classical solutions of the local GNLS action.
    
 Because the notation is more compact, we temporarily switch to the
antihermitean matrix notation of Eq.~(\ref{pot}).
The local GNLS model, normalized appropriately,   has the action
\cite{corn74}:
\begin{equation}
\label{gnlsm}  
I_{GNLS}= \frac{-M}{g^2}\int d^3xTr[U^{-1}D_iU]^2
\end{equation}
where $U$ is a unitary matrix transforming as $U\rightarrow VU$ under the
gauge transformation
\begin{equation}
\label{gtrans}
A_i\rightarrow VA_iV^{-1}+V\partial_iV^{-1}.
\end{equation}
The classical equations for $U$ express this quantity in terms of the $A_i$
\cite{corn74}, with the result
\begin{equation}
\label{upert}
U=e^{\omega};\;\;\omega = \frac{-1}{\nabla^2}\partial \cdot A
+\frac{1}{\nabla^2}\left \{[A_i,\partial
_i\frac{1}{\nabla^2}\partial \cdot A]+\frac{1}{2}[\partial \cdot
A,\frac{1}
{\nabla^2}\partial \cdot A]+\cdots \right \}
\end{equation}
showing the appearance  of  massless scalars.
More generally, since $U^{-1}D_iU$ is a gauge transformation of $A_i$,
functional integration over the $U$ is equivalent to projecting out the
gauge-invariant part of the mass term \cite{corn87,kk}.  Note that the 
term linear in $A_i$ of  the GNLS model field $U^{-1}D_iU$ is 
the transverse part of $A_i$.  This linear term is Abelian, and all
higher-order terms of $\omega$ in Eq.~(\ref{upert}) are non-Abelian.

  [Greensite and
Olejnik \cite{greenole} have conjectured that in certain instances
operators such
as $\nabla^{-2}$ should be replaced by $D^{-2}$, where $D_i=\partial_i+A_i$
is the covariant derivative.  Their lattice calculations  show that
$D^{-2}$ is a finite-range operator, with no massless poles; this is
reasonable, because it contains  gauge-potential condensate terms, but it
is not obvious where the long-range pure-gauge excitations responsible for
confinement, such as we have in Eq.~(\ref{upert}), are.  We will not follow
this line of reasoning here.]

It is now straightforward, if tedious, to verify that the two- and
three-point terms of the non-local GNLS action give rise precisely to  (the
leading mass terms of) the two-point function $S_2$ and the pinch/gauge
technique three-point function of Eq.~(\ref{3ptfunct}).  Moreover, the GNLS
action integrated over $U$ automatically satisfies the Ward identities to
all orders, just because it is the solution of the FSE for a gauge-invariant
Hamiltonian.  

\subsection{The second-leading term}

We already know that the next-leading term, of $\mathcal{O}(1/M)$, in the
expansion of the  two-point function $S_2$ is the conventional Abelian
action involving $F_{ij}^2$.  It is obvious without any calculation that
the Abelian action will, at a minimum, be gauge-completed to the full
Yang-Mills action with its three- and four-point vertices.  These come from
the three- and four-point functions in the expansion of $S$ as given in
Eq.~(\ref{nonabel1}).  The desired terms of the Yang-Mills action are
straightforwardly found either by direct solution of the FSE or from the
dressed-loop expansion, which always contain all the terms of the action of
the underlying theory divided by some sum of two-point functions $\Omega$. 
For example, we show in the Appendix that the three-point function has the
term
\begin{equation}
\label{3pvertex}
\Omega_{ijk}^{abc}(k_1,k_2,k_3)=[\Omega (1)+\Omega (2)
+\Omega (3)]^{-1}f^{abc} [\delta_{ij}(k_1-k_2)_k+c.p.] +\dots
\end{equation} 
where the term in square brackets is the free Yang-Mills three-point vertex
and each $\Omega (i)$ is replaced by $M$ to find the leading term in the
mass expansion.  There is a plethora of other terms, which either cancel
among themselves or give total divergences.   Of course, higher-order
gauge-invariant terms
may arise from higher-order coefficient functions in the gauge-potential
expansion of $S$, Eq.~(\ref{revgauge}), but we will not consider them,
since they are necessarily accompanied by higher powers of $1/M$. 

In the Abelian case the $\mathcal{O}(1/M)$ term is of the correct functional
form, but with a coefficient twice as small as it should be, and the same
problem arises for the non-Abelian case.  This results in a gauge mass of
$\sqrt{2}M$ instead of $M$, as pointed out in Sec.~\ref{abelian}.  In the
next section we consider a modification of the straightforward mass
expansion of the type of Eq.~(\ref{replace}) that forces the correct mass.

\section{\label{massexpand} The final conjecture and its consequences }

\subsection{Heuristic mass expansion}

What we have so far in the gauge-completed mass expansion to second order is
the sum of a GNLS and a Yang-Mills term, but with the wrong mass.
 What we need is an approximation to this two-term action that has the
correct mass, in part because solitons are
described in this momentum range and decay at a rate $\sim \exp (-M\rho )$. 
In any event, it is clear that the first two
terms in any sensible infrared expansion consist first of a gauge-invariant
mass term and second of a standard Yang-Mills action.   

Rather than stick to a strict expansion in powers of $\nabla^2/M^2$, 
we conjecture that, as in the Abelian case, we can replace
  $\Omega =
\sqrt{-\nabla^2+M^2}$ by a leading term $(Z/M)(-\nabla^2+M^2)$, where $Z$ is
a coefficient of $\mathcal{O}(1)$.

The mathematical motivation for least-square  fits of operators is
well-known.  Consider a normal
operator $P$, expressed in terms of its eigenvalues and eigenfunctions:
\begin{equation}
\label{pop}
P=\sum|n\rangle \lambda_n\langle n |.
\end{equation}
Any function of $P$, call it $f(P)$, is expressed by replacing $\lambda_n$
by $f(\lambda_n)$.  With the operator norm $Tr P^{\dagger}P$, we define a
relative RMS distance between two operators $f(P)$ and $g(P)$ by:
\begin{equation}
\label{rmsop}
\left\{ \frac{Tr[f(P)-g(P)][f(P)-g(P)]^{\dagger}}{Trf(P)f(P)^{\dagger}}
\right\}^{1/2}=\left\{\frac{\int d\lambda \rho (\lambda ) |f(\lambda
)-g(\lambda )|^2}
{\int d\lambda \rho (\lambda )|f(\lambda )|^2}\right\}^{1/2},
\end{equation}
where
\begin{equation}
\label{evdensity}
\rho (\lambda )=\sum \delta (\lambda -\lambda_n)
\end{equation}
is the density of eigenvalues.  One could also modify this density by
multiplying it by a non-negative function $q(\lambda )$ to emphasize a
certain range of eigenvalues, so that the weight in the integral is $\rho
(\lambda )q(\lambda )$.  

The eigenvalues of $P=-\nabla^2$ are the squared momenta $k^2$, positive for
real $k$.  We really want our approximation of $\Omega$ to be fairly good
for imaginary $k$, so the above discussion is not very useful.  Moreover,
the operators involved are not in trace class, so divergences arise. 
Instead, we take a rather simpleminded point of view, asking what is the
best fit, in the least-squares sense, of the function $Z(1-x^2)$ to the
function $\sqrt{1-x^2}$ in the interval $0\leq x \leq 1$.  Here $x^2$
represents $\nabla^2/M^2$, and positive values for this operator suggest
that we are applying it to a special class of functions representable by
Laplace transformation, with the Laplace-transformation weight peaked
around $M$.  This is indeed the property of the functions that enter into
FSE center vortices, as exemplified in  the Abelian center vortex of
Eq.~(\ref{bfield}).

For a uniform weight over $0\leq x\leq 1$ we find the normalized
least-squares integral $I_{lx}(Z)$:
\begin{equation} 
\label{rms}
I_{ls}(Z)=\left
[\frac{\int_0^1dx[Z(1-x^2)-\sqrt{1-x^2}]^2}{\int_0^1dx(1-x^2)}
\right ]^{1/2}.
\end{equation}
Minimizing on $Z$ gives $Z=\frac{45\pi}{128}\simeq 1.10$, and the minimum
value of $I_{ls}$ is about 0.22.  If we replace $x^2$ by $x$ in the
integrand of Eq.~(\ref{rms}), which corresponds to a different weight, we
get $Z=1.2$.  Both values are near unity, as expected, and the value
$I_{ls}\simeq 0.22$ suggests the relative accuracy of this least-squares
fit.  

\subsection{The final conjecture:  Relating $d$- and $d-1$-dimensional
dynamics}

 The final form of the conjecture, expressed in terms of the $d=4$ variables
$g^2,M$ is then:
\begin{equation}
\label{finalconj}
-2S= -I_{d=3}\rightarrow  \frac{2MZ}{g^2}\int
d^3xTr[U^{-1}D_iU]^2\}+\frac{Z}{Mg^2}
\int d^3x TrG_{ij}^2+\mathcal{O}(M^{-3}).
\end{equation}

We now compare this to the canonical $d=3$ form of the conjectured action,
action, which is:
\begin{equation}
\label{cand3}  I_{d=3}=-\int
d^3x\left
\{\frac{1}{2g_3^2}TrG_{ij}^2+\frac{m_3^2}{g_3^2}Tr[U^{-1}D_iU]^2\right \}.
\end{equation}
Here $G_{ij}$ is the non-Abelian field strength, $D_i=\partial_i+A_i$ is the
covariant derivative, and the unitary matrix $U$ is the GNLS
field, as before; the gluon mass is $m_3$ and the $d=3$ coupling, with
dimensions of mass, is $g_3^2$.  Equating $I_{d=3}$ with $2S$ leads to:
\begin{equation}
\label{compare}
\frac{ZM}{g^2}=\frac{m_3^2}{2g_3^2};\;\;g^2=\frac{2Zg_3^2}{M}.
\end{equation}
These equations yield $m_3=M$, as expected, plus
\begin{equation}
\label{compare2}
g^2=\frac{2Zg_3^2}{m_3}.
\end{equation}
(Note that the $d=3$ quantities scale properly at large $N$ if their $d=4$
counterparts do.)  Presumably the $d=4$ coupling $g^2$ that occurs in these
formulas is actually the running coupling $g^2(M^2)$ evaluated at the gluon
mass scale.

We can now make an estimate of a pure $d=4$ quantity in terms of a pure
$d=3$ quantity, from Eq.~(\ref{compare2}) and earlier $d=3$ results.  In
$d=3$ one quantity of particular interest  is the
dimensionless  ratio $m_3/g_3^2$.  This ratio has been estimated in a
number of continuum and lattice studies
\cite{chk,cornyan,alexn,buchphil,corn1998,eber,karsch,hkr,ckp}, and we can
see whether this $d=3$  dynamical quantity can correctly predict the  
running coupling $g^2(M^2)$ at the gluon mass scale.   Or
we can reverse the problem and use estimates of the running coupling to
predict $m_3/g_3^2$. 
There is no particular reason to think that the dynamics of the
action defined by the exact vacuum wave functional, before truncation to two
terms of a mass expansion, should be precisely that
of $d=3$ 
QCD.  Nonetheless, if our conjecture is to be believed there
should not be gross discrepancies.  

In $SU(2)$ gauge theory various authors 
\cite{chk,cornyan,alexn,buchphil,corn1998,eber,karsch,hkr,ckp} give a value
$m_3/g_3^2\simeq 0.32$, and one $SU(3)$ lattice study \cite{naka} gives a
value of 0.48.  The quantity $m_3/g_3^2$ should be linear in $N$ of $SU(N)$
for large $N$, and the factor 3/2 nicely converts the $SU(2)$ values to the
$SU(3)$ value, so we use 0.48 as the $SU(3)$ value.   We then find a value
for the strong coupling (with no quarks) $\alpha_s(M^2)\simeq 0.33Z$ that is
in fairly
good agreement with the one-dressed-loop approximation found in the
original paper on dynamical gluon mass generation \cite{corn82}.
 This paper gives a
one-dressed-loop equation for the running charge with dynamical gluon mass
generation.   At the momentum scale of the gluon mass $M$:
\begin{equation}
\label{gest}
\alpha_s(M^2)=\frac{g^2}{4\pi}=\frac{12\pi}{[11N-2N_f]\ln
[5M^2/\Lambda^2)]}\simeq 0.4,
\end{equation}
where the numerical value is based on the estimates $M=0.6$ GeV,
$\Lambda=0.3$ GeV, and the absence of quarks ($N_f=0$).  Of course, these
numbers for $M$ and $\Lambda$ are themselves uncertain, if only because
Eq.~(\ref{gest}) is a one-dressed-loop equation.  

According to this one-dressed-loop equation, accounting for three light
flavors multiplies the no-quark value by $11/9\simeq 1.2$.  If we assume
that this correction applies to the FSE result of this paper, which as it
stands does not account for quarks, our estimate of $\alpha_s(M^2)$
increases to about $0.4Z$.

Several papers have extracted values of $\alpha_s(0)\simeq 0.7\pm 0.3$ from
various scattering data sensitive to low-momentum effects \cite{natale}
that could diverge if there were no gluon mass.  The three-quark value that
we give of 0.4$Z$ is a little smaller, but in quite reasonable agreement
considering the approximation that is inherent in a two-term
truncation of the FSE exponent $S$ and our lack of knowledge of $Z$..  

It has been argued \cite{nair} that $m_3/g_3^2$ for $SU(N)$ is very closely
approximated
by the simple analytic function
\begin{equation}
\label{nair}
\frac{m_3}{g_3^2}=\frac{N}{2\pi};
\end{equation}
the present author \cite{corn82} has argued for a ratio that should be
fairly close
to $15N/(32\pi )$, which differs from the above by only a few percent.  One
then has a simple analytic formula for $\alpha_s(0)$.
Using the value from Eq.~(\ref{nair}) in Eq.~(\ref{compare2})
yields the
amusing, if not very accurate, formula
\begin{equation}
\label{compare3}
\alpha_s(M^2)=\frac{Z}{N}\simeq \frac{1}{N}.
\end{equation}

We can play the same game in one less dimension for the $d=2+1$ FSE,
 beginning with an exponent $S$ for the wave functional that is the
trivial dimensional reduction of what we began with in $d=3+1$.  The result
is a $d=2$ action with, as in $d=3+1$,  a mass term and a kinetic term. 
This is not the standard
$d=2$ QCD action, which is a free field theory.   We compare this to a
conjecture made long ago by
Greensite \cite{green79}, arguing that $S$ for the FSE  was just the
usual Yang-Mills action in one less dimension.  Unfortunately, as
\cite{greenole} notes, if Greensite's 1979 conjecture is applied in $d=2+1$,
the effective action is the familiar $d=2$ free-field QCD, which would lead
to confinement of all representations of $SU(N)$, not just those with
$N$-ality nonzero.  This is not the right behavior for $d=2+1$.  
But in our case once again the action is the Yang-Mills term plus a
GNLS model mass term; this action has \cite{corn98} center vortices; they
are point-like objects in $d=2$.  A condensate of these
solitons leads  to confinement, but only of group representations that have
$N$-ality $\not\equiv$ 0 mod $N$; other representations (such as the
adjoint) are blind to the long-range parts of center vortices.  This is the
correct behavior for $d=2+1$ gauge theories.  However, if the mass term
were not present in $I_{d=2}$ this action, which is supposed to carry all
the information about $d=2+1$ gauge theories, would reduce to the standard
Yang-Mills action in $d=2$.  The conventional treatment of $d=2$ gauge
theories, which (in the absence of dynamical matter fields) are free field
theories, finds confinement through the long-range free gluon propagator,
and all representations are confined.  But with the mass term the gluon
propagator is short-ranged and confinement comes from the pure gauge
long-range parts of center vortices. 

It is far from trivial to calculate the properties of the center-vortex
condensate in $d=2$, and so we cannot relate the $d=3$ coupling to the
string tension that would be found from the $d=2$ effective action.

\section{\label{conclusions} Summary and conclusions}

We have conjectured that to a reasonable approximation the dominant
quasi-infrared part of the vacuum wave
functionals for the $d=3+1$ and $d=2+1$ FSE are actions in one less
dimension consisting of a  
Yang-Mills term and a GNLS model term, showing  gauge-invariant dynamical
mass generation.  Two
main conclusions follow:
\begin{enumerate}
\item Given the usual entropy-dominance argument, these wave functionals
show confinement through center vortices, such
that only group representations with $N$-ality $\not\equiv 0$ mod $N$ are
confined.
\item In $d=3+1$ we can appeal to earlier works estimating the ratio
$m_3/g_3^2$ in the $d=3$ action of the FSE to make the estimate
$\alpha_s(M^2)\simeq 0.4Z$, where $Z$ is a renormalization constant that we
have very crudely estimated to be in the neighborhood of 1.1-1.2.  This can
be compared to an earlier estimate, based on the original work on dynamical
gluon mass generation, of $\alpha_s(M^2)\simeq 0.4$.  Both these estimates
have three light flavors of quarks.   This is to be compared to
compared to phenomenological estimates \cite{natale}, also with three light
quarks,  of $\alpha_s(0)\simeq 0.7\pm 0.3$.   
\end{enumerate}

It would be interesting to verify this structure of the FSE vacuum wave
functionals through lattice simulations.  

\acknowledgments

I am happy to acknowledge valuable conversations with \u{S}tefan Olen\'{\i}k
about Ref. \cite{greenole}.

\newpage

\appendix

\section{\label{review} A brief review of the FSE and solution methods}

The purpose of this review of known material \cite{corn87} is to indicate
the plausibility
of constructing an infrared-accurate and gauge-invariant form of the wave
functional $\psi$, based on a single operator $\Omega$, obeying a
non-linear Schwinger-Dyson equation.  In ordinary quantum mechanics this is
exactly what happens except that the ``Schwinger-Dyson equation" is simply
algebraic.  The FSE for scalar field theories is really nothing but
ordinary quantum mechanics for infinitely-many coupled oscillators, so we
review it and its connection to quantum mechanics, then go on to gauge
theories.

\subsection{The Schr\"odinger equation}

The general principles of solving the FSE in terms of an
operator $\Omega$ are most easily understood from the ordinary
Schr\"odinger equation. Consider the quadratic/quartic Hamiltonian
\begin{equation}
\label{phi4}
H=\frac{-1}{2}(\frac{d}{dx})^2+\frac{1}{2}\omega^2x^2+\frac{1}{4!}\lambda
x^4.
\end{equation}
The ground-state solution is $\psi =\exp (-S)$, with
\begin{equation}
\label{groundstate}
S=\frac{1}{2}\Omega x^2+\frac{1}{4!}\Omega_4x^4+\dots
\end{equation}
Following \cite{corn87}, we substitute $\psi$ in the Schr\"odinger equation
saving only terms through $\Omega_6$  and find:
\begin{equation}
\label{phi4se}
\Omega_6=\frac{-5\Omega_4^2}{3\Omega};\;\;
\Omega_4=\frac{\lambda}{4\Omega}+\frac{\Omega_6}{8\Omega};\;\;
\Omega^2=\omega^2+
\frac{1}{4}\Omega_4;\;\; E=\frac{1}{2}\Omega. 
\end{equation}
It is easy to solve the equation for $\Omega_4$ to derive a quartic
equation for $\Omega$.  One can go on to any order this way, expressing (in
principle, at least) every $n$-point coefficient up to a given highest value
of  $n$
 in terms of $\Omega$, and ending up with a
non-linear dressed-loop equation for $\Omega$.   Consider  now the case
$\omega =0$, for which the perturbative expansion coefficient $\lambda
/\omega^3$ diverges.  Then through the six-point term we find the expression
$E =(3\lambda
/272)^{1/3}$, which has the value 0.2226$\lambda^{1/3}$. This is within a
few percent of the numerical answer of $0.2311\lambda^{1/3}$.

\subsection{Field theories other than gauge theories}

  For
simplicity of exposition, we begin  with a scalar field theory.  Take
the FSE Hamiltonian to be
\begin{equation}
\label{revh}
H=\int d^3x\left [\frac{-1}{2}(\frac{\delta }{\delta
\phi})^2+\frac{1}{2}(\nabla
\phi^2)+\frac{1}{2}m^2\phi^2+V(\phi )\right ].
\end{equation}
where the potential $V$ contains cubic and higher terms.  Ref. \cite{corn87}
showed that the vacuum wave functional could be expressed as a $d=4$
partition function:
\begin{equation}
\label{revfinal}
e^{-S}=const. \times \int (d\Phi ) \exp [-I_0(\Phi )-I_0(\hat{\phi}_0)
-\int V(\Phi + \hat{\phi}_0)].
\end{equation}
In this partition function, space-time integrals are of the form of an
integral over a Euclidean time $\tau$ and all of three space:
\begin{equation}
\label{revint}
 \int_0^{\infty}d\tau\int d^3x.
\end{equation}
The argument of $S$ is the field $\phi (x)$, and the field
$\hat{\phi}(x)$ depends on $x=(\tau ,x)$ as:
\begin{equation}
\label{phihat}
\hat{\phi}_0(x)=e^{-\Omega_0 \tau}\phi(x)
\end{equation}
with
\begin{equation}
\label{defom0}
\Omega_0=\sqrt{M_0^2-\nabla^2}.
\end{equation}
The free action $I_0$ is:
\begin{equation}
\label{freeact}
I_0(\Phi )= \frac{1}{2}\int [(\partial_{\tau}\Phi)^2+(\nabla \Phi )^2]
\end{equation}
and the inverse of the free-action operator is the free propagator
\begin{equation}
\label{freeprop}
\Delta_0 =
\langle x|\frac{1}{2\Omega_0}[e^{-\Omega_0|\tau -\tau'|}-
e^{-\Omega_0(\tau + \tau')}]|x'\rangle.
\end{equation}
The first term in the propagator is
the usual Euclidean propagator:
\begin{equation}
\label{revprop2}
\langle x|\frac{1}{2\Omega_0}e^{-\Omega_0|\tau
-\tau'|}|x'\rangle=\frac{1}{(2\pi )^4}\int d^4k\frac{e^{ik\cdot
(x-x')}}
{k^2+m^2}.
\end{equation}
For purposes of calculating the energy eigenvalue, this is the only term
that needs to be saved in $\Delta_0$, but the second term of the propagator
in  Eq.~(\ref{freeprop}) is necessary for calculating the wave functional.
 
Either by working out the partition function of Eq.~(\ref{revfinal}) or by
direct substitution in the Schr\"odinger equation one sees that     the
vacuum functional $\psi$ has the general form (using a streamlined but
transparent  notation):
\begin{equation}
\label{revvac}
\psi =
e^{-S};\;\;S=\frac{1}{2}\int \int \phi\Omega\phi+\sum_N\frac{1}{N!}\int
\dots \int \Omega_N\phi_1
\dots \phi_N.
\end{equation}
For purely three-dimensional equations, such as this, the unadorned integral
sign simply  indicates $\int d^3x$, where $x$ is the argument of a
corresponding $\phi$ (and a sum
over discrete indices, if any), with $\Omega$ and the $\Omega_N,N\geq 3$,
as translationally-invariant form factors in the arguments of the $\phi$. 
The partition function form in Eq.~(\ref{revfinal} can be addressed with
the  well-known resummation techniques
\cite{cjt} of the dressed-loop expansion.  The effect of these rules is to
 remove a large fraction of one-particle-reducible graphs, as required for
the dressed-loop expansion.   In part, this amounts to a general replacement
(but not quite everywhere) of the free operator $\Omega_0$ by a dressed
operator $\Omega$ that satisfies a non-linear Schwinger-Dyson equation. 
This operator is precisely the same as the $\Omega$ that occurs in the
quadratic term of the wave functional in Eq.~(\ref{revvac}). 

For further details of this formalism for scalar field theories, see 
\cite{ccds} which uses it for calculating some terms in the Wigner
distribution function.

\subsection{Gauge theories}

For gauge theories the same general structure holds; the principal problem
remaining is to enforce gauge invariance.  
The canonical momentum and Hamiltonian are  represented by
\begin{equation}
\label{revcan}
\Pi_i^a \rightarrow -ig^2\frac{\delta}{\delta A_i^a};
H=\int d^3x [-\frac{1}{2}g^2(\frac{\delta}{\delta A_i^a})^2+\frac{1}{2g^2}
(B_i^a)^2]
\end{equation}
where $B_i^a$ is the chromomagnetic field strength.
The generator of infinitesimal gauge transformations is $D_j^{ab}(-i\delta
/\delta A_j^b)$, and this must annihilate $\psi$.  The exponent $S$ in
$\psi$ has the form given in Eq.~(\ref{nonabel1}), repeated here for
convenience:
\begin{equation}
\label{revgauge}
g^2S=\frac{1}{2!}\int \int A_i^a\Omega_{ij}A_j^a+\frac{1}{3!}\int \int \int
A_i^aA_j^bA_k^c
\Omega_{ijk}^{abc}+\dots
\end{equation}
Invariance of $S$ under infinitesimal gauge transformations is trivial for
the two-point function $\Omega_{ij}$; this quantity must be conserved, so
that in Fourier space
\begin{equation}
\label{revcons}
\Omega_{ij}(k)=\Omega (k)P_{ij}(k);\;\;
P_{ij}=\delta_{ij}-\frac{k_ik_j}{k^2}.
\end{equation}
   For the free theory $\Omega_0(k)=k$.  

Gauge invariance is more complicated for higher-point functions. 
Annihilating $\psi$ with the generator of gauge transformations yields a
set of {\em ghost-free} Ward identities (these Ward identities also apply
to the pinch technique \cite{corn82,binpap} construction of gauge-invariant
Green's functions).  For example, the Ward identity for the three-point
function is:
\begin{equation}
\label{revward}
k_{1i}\Omega_{ijk}^{abc}(k_1,k_2,k_3)=f^{abc}
[\Omega_{jk}(2)-\Omega_{jk}(3)]
\end{equation}
where $\Omega_{jk}(2)\equiv \Omega_{jk}(k_2)$, etc. 

Now turn to the FSE itself.  The equation determining the three-point
function has the general form
\begin{equation}
\label{rev3pt}
\Omega_{il}(1)\Omega_{ljk}^{abc}+\Omega_{jl}(2)\Omega_{lik}^{bac}
+\Omega_{kl}(3)\Omega_{lij}^{cab}=f^{abc}\Gamma_{ijk}.
\end{equation}
The right-hand side $\Gamma_{ijk}$ comes from the cubic term in $H$, plus
another term from the five-point function.  The Ward identity for
$\Gamma_{ijk}$ is determined by the above equation plus the Ward identities
for the two- and three-point functions as already given, and multiplying
both sides of Eq.~(\ref{rev3pt}) by $k_{1i}$ yields:
\begin{equation}
\label{revgamma}
k_{1i}\Gamma_{ijk}=\Omega_{jk}^2(3)-\Omega_{jk}^2(2).
\end{equation} 
For free particles, with $\Omega = \Omega_0$, this is satisfied by the usual
free three-point vertex
\begin{equation}
\label{rev3ptfree}
\Gamma^0_{ijk}=i(k_1-k_2)_k\delta_{ij}+ c.p.
\end{equation}

The reader can verify that the FSE equation (\ref{rev3pt}) has a solution of
the form:
\begin{equation}
\label{revfse}
\Omega_{ijk}^{abc}(k_1,k_2,k_3)=[\Omega (1)+\Omega (2)
+\Omega (3)]^{-1}f^{abc}\left \{\Gamma_{ijk}+\{\Omega
(1)\frac{k_{1i}}{k_1^2}
[\Omega_{jk}(2)-\Omega_{jk}(3)]+c.p.\}\right \}
\end{equation}
which respects the Ward identity of Eq.~(\ref{revward}), by virtue of the
massless pole terms of Eq.~(\ref{revfse}).  It should now be clear that
these longitudinally-coupled massless excitations will occur, as a result
of enforcing gauge invariance, for every n-point function.  We will shortly
identify these with couplings of the GNLS field
introduced in our conjecture for the infrared-effective action.

So far the vertex function $\Gamma_{ijk}$ is undetermined.  As \cite{corn87}
argues, one can carry out a program of expressing all higher-point
functions in terms of the two-point function, and then the FSE (or the
equivalent dressed-loop expansion) becomes a non-linear, non-perturbative
equation for this two-point function $\Omega$.   The idea, known also as
the gauge technique, is to find an infrared-effective approximation to
$\Gamma_{ijk}$ that exactly satisfies  the Ward identity (\ref{revgamma})
for any $\Omega$.  One can, at least in principle, find such
infrared-effective approximations for four- and higher-point functions as
functionals of $\Omega$.  In fact, a very general form for the ``solution"
to the Ward identity for the three- and four-point functions is known
\cite{cornhou,papa} for arbitrary dependence of $\Omega$ on momentum.   The
word ``solution" is enclosed in quotes because it is not unique; any
completely-conserved term can be added to the ``solution" for
$\Gamma_{ijk}$, for example.  But the point is that purely-conserved terms
are of higher order in momenta than the terms saved in the gauge technique.

The general solution of \cite{cornhou} is:
\begin{equation}
\label{cornhoueq}
\Gamma_{ijk}=\delta_{ij}(k_1-k_2)_k-\frac{k_{1i}k_{2j}}{2k_1^2k_2^2}
(k_1-k_2)_l\Pi_{lk}(k_3)-[P_{il}(k_1)\Pi_{lj}(k_2)
-P_{jl}(k_2)\Pi_{li}(k_1)]\frac{k_{3k}}{k_3^2}+c.p.
\end{equation}
where the first term on the right is the free vertex $\Gamma^0_{ijk}$ and
$\Pi_{ij}(k)\equiv P_{ij}(k)\Pi)k)$ is the transverse pinch-technique
\cite{corn82,binpap} self-energy, related to $\Omega_{ij}$ by:
\begin{equation}
\label{omegaeq}
\Omega_{ij}^2=P_{ij}[\Omega_0^2+\Pi \{\Omega\}]
\end{equation}
where $\Omega_0^2=k^2$ is the free gluon contribution.

In the simple case studied by us, $\Pi=M^2$, and the resulting expression
for $\Gamma_{ijk}$ is:
\begin{equation}
\label{3ptward}
\Gamma_{ijk}=\delta_{ij}(k_1-k_2)_k+\frac{M^2}{2}\frac{k_{1i}k_{2j}(k_1-k_2)_k}
{k_1^2k_2^2}+c.p.
\end{equation}

As  we saw above, in ordinary quantum mechanics in one spatial dimension $x$
the exponent $S(x)$ of $\psi$ can be determined systematically from a set
of non-linear algebraic equations, such that each term of
$\mathcal{O}(x^3)$ or higher can be expressed in terms of the quadratic
coefficient $\Omega$.  Finally, $\Omega$ is determined by a single
non-linear equation, equivalent to a dressed-loop expansion.  Combining the
pinch technique and the gauge technique gives  a
completely analogous program for gauge theories, based on ``solving" the
Ward identities insuring gauge invariance.  While this program can only be
carried out approximately, it is  gauge-invariant by construction. 
Ultimately it yields a dressed-loop equation for a single transverse
operator $\Omega_{ij} (k)\equiv P_{ij}(k)\Omega(k)$.   The
pinch-technique self-energy $\Pi$ is itself a complicated function of
$\Omega$,  found by using dressed propagators of the general form given in
Eq.~(\ref{freeprop}), with $\Omega_0$ replaced by $\Omega$ and with
appropriate vector kinematics.   In effect, $\Pi$ is the on-shell
self-energy and any $\pm i\Omega (k)$ occurring in $\Pi$  is a fourth
component of a Euclidean four-vector $(k_4,k)$ that is on-shell, by
which we mean that   $k_4^2+k^2+M^2=0$, or $k_4=\pm i\sqrt{k^2+M^2}$. 

All that we need from this development in the main text is
Eq.~(\ref{3ptward}), which will be used in the large-$M$ expansion of the
three-point function $\Omega^{abc}_{ijk}$.

\end{document}